\documentclass[a4paper,fleqn,usenatbib]{mnras}
\usepackage[T1]{fontenc}
\usepackage{ae,aecompl}

\usepackage{graphicx}	
\usepackage{amsmath}	
\usepackage{amssymb}	
\usepackage{rotfloat} 
\usepackage{graphicx}
\usepackage{longtable}
\usepackage{url}
\usepackage{dcolumn}
\usepackage{placeins}
\usepackage{multirow} 
\usepackage{hyperref}
\usepackage{rotating} 
\usepackage{bm}		
\usepackage{pdflscape}	
\usepackage{ragged2e}
\justifying
\usepackage[T1]{fontenc}
\usepackage[none]{hyphenat}
\usepackage{color,soul}
\usepackage{adjustbox}
\usepackage{multicol}
\usepackage{caption}
\usepackage{subcaption}
\usepackage[english]{babel}
\usepackage{blindtext}
\usepackage{tabu}
\usepackage{booktabs,caption,fixltx2e}
\usepackage[flushleft]{threeparttable}
\usepackage{array}
\usepackage{afterpage}
\usepackage{capt-of}
\usepackage[stable]{footmisc} 
\usepackage{booktabs}

\newcommand{\lya}{Ly$\alpha$}
\newcommand{\lyb}{Ly$\beta$}

\newcommand{\HI}{\mbox{H\,{\sc i}}}

\newcommand{\HeII}{\mbox{He\,{\sc ii}}}

\newcommand{\OII}{\mbox{O\,{\sc ii}}}
\newcommand{\OIII}{\mbox{O\,{\sc iii}}}
\newcommand{\OIV}{\mbox{O\,{\sc iv}}}
\newcommand{\OV}{\mbox{O\,{\sc v}}}
\newcommand{\OVI}{\mbox{O\,{\sc vi}}}
  
\newcommand{\OVII}{\mbox{O\,{\sc vii}}}
\newcommand{\OVIII}{\mbox{O\,{\sc viii}}} 

\newcommand{\CIII}{\mbox{C\,{\sc iii}}}

\newcommand{\NII}{\mbox{N\,{\sc ii}}}   
\newcommand{\NIII}{\mbox{N\,{\sc iii}}}     
\newcommand{\NIV}{\mbox{N\,{\sc iv}}}

\newcommand{\NeVII}{\mbox{Ne\,{\sc vii}}}   
\newcommand{\NeVIII}{\mbox{Ne\,{\sc viii}}}   
\newcommand{\neviii}{\mbox{\tiny Ne\,{\sc viii}}}   
\newcommand{\NeIX}{\mbox{Ne\,{\sc ix}}}   

\newcommand{\zabs}{$z_{\rm abs}$}

\newcommand{\vrel}{$v_{\rm rel}$}

\newcommand{\be}{\begin{equation}}
\newcommand{\en}{\end{equation}}

\def\lam{$\lambda$}

\def\kms{km~s$^{-1}$}

\def\cmsq{cm$^{-2}$}
\def\cmcb{cm$^{-3}$}

\newcommand{\CLOUDY}{\mbox{\scriptsize{CLOUDY}}}

\newcommand{\pirate}{$\Gamma_{\rm HI}$}
\newcommand{\fesc}{$f_{\rm esc}$}
\newcommand{\nh}{$n_{\rm \tiny H}$}

\newcommand{\fhi}{$f_{\rm \tiny HI}$}
\title[Ionization mechanisms of \NeVIII\ absorbers]{Implications of an updated ultraviolet background for the ionization mechanisms of intervening \NeVIII\ absorbers}
\author[Hussain, T. et al.]
{
\parbox[t]{\textwidth}{ 
Tanvir Hussain$^{1,2}$\thanks{E-mail: \href{mailto:tanvir@iucaa.in}{tanvir@iucaa.in}},  
Vikram Khaire$^{3}$\thanks{E-mail: \href{mailto:kvikram@ncra.tifr.res.in}{kvikram@ncra.tifr.res.in}}, 
Raghunathan Srianand$^{1}$,
Sowgat Muzahid$^{4}$ and 
Amit Pathak$^{2}$
} 
\vspace*{10pt}\\ 
$^{1}$ Inter-University Centre for Astronomy and Astrophysics, Post Bag 4, 
Ganeshkhind, Pune 411\,007, India \\ 
$^{2}$ Department of Physics, Tezpur University, Tezpur 784\,028, India\\ 
$^{3}$ National Centre for Radio Astrophysics, Tata Institute of Fundamental Research, Pune 411\,007, India\\
$^{4}$ Leiden Observatory, University of Leiden, P.O. Box 9513, 2300 RA Leiden, the Netherlands \\  
}   

\date{Accepted\underline{\hspace{1cm}}. Received\underline{\hspace{1cm}}; in original form\underline{\hspace{1cm}}}

\pubyear{2016}

\begin{document}
\label{firstpage}
\pagerange{\pageref{firstpage}--\pageref{lastpage}}
\maketitle

\begin{abstract}
\NeVIII\ absorbers seen in QSO spectra are useful tracers of warm ionized gas, when collisional ionization is the dominant ionization process. While photoionization by the ultraviolet background (UVB) is a viable option, it tends to predict large line-of-sight thickness for the absorbing gas. Here, we study the implications of the recently updated UVB at low-$z$ to understand the ionization mechanisms of intervening \NeVIII\ absorbers. With the updated UVB, one typically needs higher density and metallicity to reproduce the observed ionic column densities under photoionization. Both reduce the inferred line-of-sight thicknesses of the absorbers. We find a critical density of $\geqslant5\times10^{-5}$~\cmcb\ above which the observed $N(\NeVIII)/N(\OVI)$ can be reproduced by pure collisional processes. If the gas is of near solar metallicity (as measured for the low ions) then the cooling timescales will be small (<$10^{8}$~yrs). Therefore, a continuous injection of heat is required in order to enhance the detectability of the collisionally ionized gas. Using photoionization models we find that in almost all \NeVIII\ systems the inferred low ion metallicity is near solar or supersolar. If we assume the \NeVIII\ phase to have similar metallicities then photoionization can reproduce the observed $N(\NeVIII)/N(\OVI)$ without the line-of-sight thickness being unreasonably large and avoids cooling issues related to the collisional ionization at these metallicities. However the indication of broad \lya\ absorption in a couple of systems, if true, suggests that the \NeVIII\ phase is distinct from the low ion phase having much lower metallicity.

\end{abstract}

\begin{keywords}
quasars: absorption lines, galaxies: intergalactic medium, cosmology: diffuse radiation
\end{keywords}
%
\section{Introduction}\label{sec1}

The early measurements of abundances of deuterium along the line-of-sight to distant QSOs \citep[]{Burles98, Omeara06,Pettini08b} 
and cosmic microwave background (CMB) observations \citep[]{Spergel03, Komatsu11} provided a good estimation of the baryonic content in the universe. A recent census of baryons at $z \lesssim 0.4$, reported by \citet[][]{Shull12b}, reveals that $28\pm11\%$ of the baryons reside in photoionized intergalactic medium (IGM) traced by the \lya\ forest \citep[]{Rauch97b,Penton04}, $18\pm4\%$ are observed in stars and cold interstellar medium (ISM) in galaxies, circumgalactic medium (CGM)  and X-ray emitting gas in clusters of galaxies and $25\pm8\%$ are found in the warm-hot intergalactic medium (WHIM) traced by broad \lya\ (BLAs) and \OVI\ absorbers. The remaining $29\pm13\%$, the so-called ``missing baryons'' are thought to reside in the hot phase of WHIM with gas densities in the range n$_{\scriptsize \rm H} \sim 10^{-6} - 10^{-5}$~\cmcb, temperatures in the range $T \simeq 10^5-10^7$ K \citep[]{persic92,Dave01} and baryonic overdensity $\Delta \sim 1-100$ \citep[]{Nevalainen15}. In comparison, \HI\ \lya\ forest absorption seen in QSO spectra at these low-$z$ originate from baryonic overdensities of $\sim1-30$ covering $N(\HI)=10^{12}-10^{14}$~\cmsq\ \citep[see Table 2 of][]{Gaikwad16}.

Since WHIM is hot and tenuous, it is best probed through X-ray absorption spectroscopy. An additional $\sim$15\% of the baryons could reside in the hot ($T>10^{6}$ K) X-ray absorbing gas traced by the \OVII\ \lam21.602 \citep[]{Nicastro05a, Nicastro05b, Nicastro08, Buote09, Fang10, Zappacosta10} and \OVIII\ \lam 18.969 absorption lines \citep[]{Fang02, Fang07}. However, due to the lack of instruments capable of  high-resolution and high S/N X-ray spectroscopy \citep[]{Yao11, Yao12, Nicastro13a, Ren14, GSantos15}, one has to rely on detecting WHIM through far ultraviolet (FUV, 900<\lam(\AA)<3000) absorption lines in the spectra of QSOs. FUV spectrographs viz., {\sl Space Telescope Imaging Spectrograph} (STIS), {\sl Cosmic Origins Spectrograph} (COS) on board the {\sl Hubble Space Telescope} ($HST$) and the {\sl Far Ultraviolet Spectroscopy Explorer} ($FUSE$) have better resolution and S/N which enable one to detect  \OVI\ \lam\lam1031,1037 \citep[e.g.][]{Tripp00, Richter04, Danforth05, Danforth08, Tripp08} and \NeVIII\ \lam\lam770,780 \citep[e.g.][]{Savage05a, Narayanan09, Narayanan11} absorption lines. Collisional ionization equilibrium (CIE) models of \citet[]{Gnat07} show that both \OVI\ and \NeVIII\ can be good tracers of the low temperature end of the WHIM having temperatures in the range $T \sim 10^{5.5}-10^{5.7}$K. 

To understand which ionizing mechanism (photoionization or collisional ionization or a combination of both) leads to the formation of such highly ionized species, it is necessary to have the ionizing ultraviolet background (UVB) radiation modeled accurately. The UVB spectrum decides (i) the derived metallicity of the gas, (ii) the density at which the observed $N(\NeVIII)/N(\OVI)$ can be produced, and (iii) the critical density above which the gas will be in pure collisional equilibrium. All these are important to decide the mechanism that keeps the \NeVIII\ phase ionized. 

The spectrum of UVB cannot be observed directly, however, can be modeled by a detailed radiative transfer of UV photons through the IGM \citep[see][]{Miralda90, Shapiro94, Haardt96, Fardal98, Shull99}. It involves using various observables as inputs such as QSO and galaxy luminosity functions \citep[]{Schmidt67, Schechter76}, spectral energy distribution (SED) of QSOs \citep[]{Telfer02}, the fraction ($f_{\rm esc}$) of \HI\ ionizing photons escaping from galaxies \citep[]{Leitherer95, Heckman01}, and the effective opacity of the IGM \citep[]{Petitjean93}. The $f_{\rm esc}$ is one of the most uncertain observables and plays a significant role in determining the spectrum of UVB \citep[see][]{Khaire13b}. At low-$z$, observations of a sample of galaxies provide a 3$\sigma$ upper limits on \fesc\ to be less than 2\% \citep[]{Cowie09, Siana10, Bridge10, Rutkowski16}. The \HI\ photoionization rate ($\Gamma_{\rm HI}$) measurements obtained using various methods \citep[see][]{Sunyaev69, Bajtlik88, Dove94a, Vogel95, Srianand96, Savaglio97} can be used to constrain the values of \fesc\ \citep[see for e.g.,][]{Inoue06, Khaire16} while modeling the UVB.

The recent studies of \HI\ column density distribution of low-$z$ \lya\ forest, using $HST$/COS data \citep[][]{Danforth16}, show that the $\Gamma_{\rm HI}$ at low redshifts (i.e., $z < 1.0$) is a factor 2 to 5 times higher \citep[]{Kollmeier14, Shull15, Wakker15, Gaikwad16b} than what is predicted by the UVB model of \citet[][hereafter, HM12]{Haardt12}. This discrepancy, pointed out as a crisis in the photon production \citep[]{Kollmeier14}, is resolved by \citet[][hereafter, KS15]{Khaire15b} using the updated QSO luminosity function \citep[i.e.,][]{Croom09, Palanque13} and star formation history of low-$z$ galaxies \citep[]{Khaire15a}. It is found that, \pirate\ obtained using only the updated QSO emissivity (i.e., with no galaxy contribution to the UVB or \fesc = 0\%), is a factor of 2 higher than that of HM12 UVB at $z<0.5$. This is also consistent with the \pirate\ measurements by \citet[]{Shull15}. With a constant \fesc\ of 4\%, the KS15 UVB provides 5 times higher \pirate\ than HM12 at $z<0.5$ consistent with the \citet[]{Kollmeier14} measurements. However, the latest \pirate\ measurements by \citet[]{Gaikwad16}, \citet{ Gurvich16}, \citet{Gaikwad16b}, and \citet{Viel16} favor results of \citet{Shull15}.
\citet[]{Gaikwad16} measured \pirate\ in four redshift bins at $z \lesssim 0.45$ using two different statistics and also provided a good accounting of associated statistical and systematic errors. Their measurements are well reproduced by  KS15 UVB with \fesc=0\% and provide a $3\sigma$ upper limit of 0.8 per cent for the \fesc\ at $z<2$. Therefore, in analysis presented in this work we use KS15 UVB with \fesc=0\%.

Using the previously published extragalactic UVBs, i.e. \citet[][hereafter, HM96]{Haardt96}, \citet[][hereafter, HM01]{Haardt01}, HM05\footnote{HM05 is the unpublished extragalactic UVB described in \citet[][]{Haardt01} and is included for use in \CLOUDY\ \citep[see][for detailed description]{Ferland13}.} and HM12 as the incident radiation, it has been found that the observed properties of \OVI\ in the low-$z$ IGM can be understood by both photoionization \citep[]{Thom08, Tripp08, Howk09, Muzahid12, Muzahid15} as well as collisional ionization models \citep[]{Danforth08}. On the other hand, for the \NeVIII\ absorbers detected so far (summarized in Table~\ref{tab1}), photoionization was not favored \citep[but see,][for an exception]{Hussain15} based on extremely large line-of-sight thickness (i.e., >1~Mpc) inferred from the photoionization models. Therefore, the previous studies of \NeVIII\ systems \citep[]{Savage05a, Narayanan09, Narayanan11, Narayanan12, Meiring13} favor these absorbers originating from a gas collisionally ionized at $T \approx 10^{5.4}-10^{5.7}$~K and thus could harbor a vast reservoir of baryons evident from their high inferred total hydrogen column densities ($N_{\rm H} \sim 10^{19} - 10^{20}$~\cmsq). Presence of BLA absorption associated to the \NeVIII\ components will favor this picture \citep[see for example,][]{Savage11}. 

In this work, using the recently updated UVB, we revisit the ionization scenarios for the \NeVIII\ absorbers at low-$z$. This article is organized as follows: in Section~\ref{sec2}, we describe the updated UVB. Discussion on the various ionization models for \NeVIII\ systems are explored in Section~\ref{sec3}. Re-analysis of the \NeVIII\ absorbers with the updated UVB and the results are given in Section~\ref{sec4}. In Section~\ref{sec5}, we summarize our conclusions. Throughout this article, we adopt an $H_0$=70~\kms Mpc$^{-1}$, $\Omega_{\rm m}$ = 0.3, and $\Omega_{\Lambda}$ = 0.7 flat cosmology \citep[]{Komatsu11}. The solar relative abundances for the heavy elements are taken from \citet{Grevesse10}.  

\section{Updated UVB \lowercase{and} \NeVIII\ systems}\label{sec2}

{%
\newcommand{\mc}[3]{\multicolumn{#1}{#2}{#3}}
\begin{table*}
\caption{Properties of intervening \NeVIII\ absorbers from the literature.}
\label{tab1}
\begin{center}
\begin{tabular}[l]{lcllllllll}\hline\hline
&  &  &  &  &  &  & Line-of-sight &  &  \\
   &  &		$\log [N(\OVI)]^{a}$   & $\log [N (\NeVIII)]^{a}$ & $\log [N_{\rm H}]^{b}$   & $\log [N(\HI)]^{c}$     & $[$X/H$]^{d}$ & thickness$^{e}$ & UVB used for  & Ref.\\ 
  \mc{1}{c}{QSO} & \zabs & (\cmsq) & (\cmsq) & (\cmsq) & (\cmsq) &  & (Mpc) & PI model  & \\ \cmidrule(lr){1-1} \cmidrule(lr){2-2} \cmidrule(lr){3-3} \cmidrule(lr){4-4}  \cmidrule(lr){5-5} \cmidrule(lr){6-6} \cmidrule(lr){7-7} \cmidrule(lr){8-8} \cmidrule(lr){9-9} \cmidrule(lr){10-10} 
HE~0226--4110  & 0.20701 & $14.38 \pm 0.01$       & $13.89 \pm 0.11$     & 20.06       & 13.87 & $- 0.5$ & $\sim$ 11 & HM96  & $[\rm AI]$ \\
3C~263	      & 0.32566 & $13.98 \pm 0.05$       & $13.99 \pm 0.11$     & 19.50       & 13.20 & $- 0.38$ &      $\sim$ 5.6  & HM96 & $[\rm AII]$   \\
PKS~0405--123  & 0.49509 & $14.39 \pm 0.01^{f}$   & $13.96 \pm 0.06^{f}$ & 19.70$^{g}$ & $\lesssim$13.50  & $- 1.0$ &   $\sim$ 2.5 & HM01  & $[\rm AIII]$\\
PG~1148+549    & 0.68381 & $14.41 \pm 0.02 $      & $13.95 \pm 0.04 $    & 19.80       & 13.60  & $- 0.5$ &   $\gtrsim$ 1 & HM96+HM01 & $[\rm AIV]$\\
PG~1148+549    & 0.70152 & $14.28 \pm 0.03 $      & $13.82 \pm 0.06 $    & 19.00     & $<$12.80 & $ > 0.2$ &   $\gtrsim$ 1 & HM96+HM01 & $[\rm AV]$ \\
PG~1148+549    & 0.72478 & $13.86 \pm 0.07 $      & $13.81 \pm 0.06 $    & 18.90     & $<$12.60 & $ > 0.0$ &   $\gtrsim$ 1 &  HM96+HM01 & $[\rm AVI]$ \\
PG~1407+265    & 0.59961 & $13.67 \pm 0.09^{h}$	 & $13.65 \pm 0.19^{h}$ & 19.65       & $<$13.40   & $0.0$    & $\sim$ 0.186$^{i}$&  HM05 & $[\rm AVII]$\\ \hline 
\end{tabular}
\begin{flushleft}
{\bf Notes:} \\
$^{a}$~Voigt profile fitted column density measured for the component at \vrel = 0~\kms, co-aligned in velocity with the low ions detected in the respective spectra.\\ 
$^{b}$~The total hydrogen column density $N_{\rm H}$ as inferred from the collisional ionization equilibrium (CIE) models. \\
$^{c}$~The neutral hydrogen column density $N(\HI)$ associated with broad \NeVIII-gas component consistent with the CIE models. \\
$^{d}$~$[$X/H$]$ ($ \equiv \log [$X/H$]_{\rm obs} - \log [$X/H$]_{\odot}$) is the logarithmic metallicity with respect to solar abundances for the low ionized species as calculated from photoionization models.\\
$^{e}$~Line-of-sight thickness as measured from photoionization models for the high ions (\NeVIII\ and \OVI).\\
$^{f}$~Measured using the apparent optical depth (AOD) method \citep[]{Savage91}.\\
$^{g}$~Computed assuming a hybrid (photoionization + collision) ionization model.\\
$^{h}$~The column density of \NeVIII\ and \OVI\ is for the component at \vrel = -- 150~\kms.\\
$^{i}$~Line-of-sight thickness of $\sim$175 kpc was reported using HM12 UVB and solar metallicity.  \\ 
$[\rm AI]$ \citet[]{Savage05a, Savage11}, $[\rm AII]$ \citet[]{Narayanan09,Narayanan12}, $[\rm AIII]$ \citet[]{Narayanan11}, $[\rm AIV, AV, AVI]$ \citet[]{Meiring13}, $[\rm AVII]$ \citet[]{Hussain15}
\end{flushleft}
\end{center}
\end{table*}
}

In this work, we study the implications of using the KS15 UVB in the photoionization models on the inferred densities, metal abundances, and line-of-sight thickness of the known \NeVIII\ absorbers. KS15 UVB uses an updated QSO emissivity and star formation rate density \citep[]{Khaire15a}. To analyse the \NeVIII\ systems, we use KS15 UVB generated for \fesc=0\% consistent with \pirate\ measurements of \citet{Shull15} and \citet[]{Gaikwad16}. 

\begin{figure} 
\centerline{
\vbox{
\centerline{\hbox{   
\includegraphics[totalheight=0.30\textheight, trim=0cm 0cm 0.75cm 0cm, clip=true]{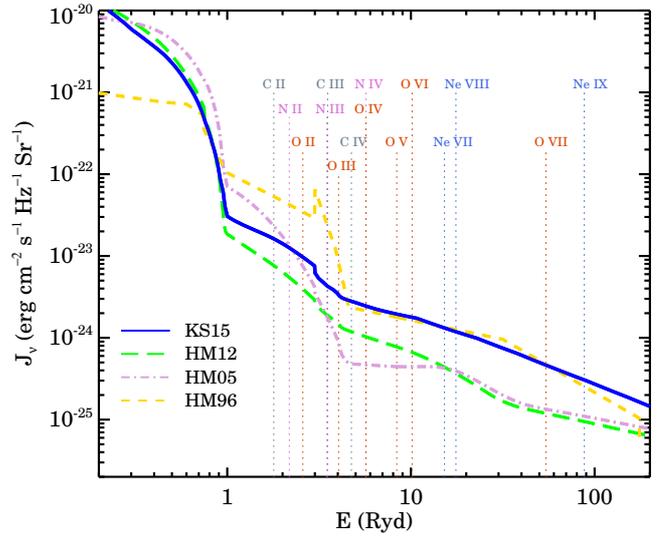}
}}
}}  
\caption{The KS15 UVB spectrum at redshift $z$ = 0.2 is shown as the blue continuous curve. For comparison, we also show the HM96, HM05 and HM12 UVBs at the same $z$ in yellow, purple and green dashed curves respectively. The vertical dotted lines (in different colors) mark the ionization energies of different ions of carbon, nitrogen, oxygen and neon as indicated.}              
\label{fig1}      
\end{figure}

In Fig.~\ref{fig1}, we plot the KS15 UVB spectrum (energy E vs specific intensity $J_{\nu}$) for \fesc=0\% at $z$ = 0.2. For comparison, we also show the HM96, HM05 and HM12 UVBs at that redshift. The dotted vertical lines mark the ionization energies of different ions of carbon, nitrogen, oxygen and neon. Clearly from this figure,  we see that the ionization energies of highly ionized species (such as \OV, \OVI, \OVII, \NeVII, \NeVIII\ and \NeIX ) fall beyond the \HeII\ ionization energy, i.e., $\rm E > 4$~Ryd.  At these energies, since  galaxies do not show any emission, the UVB is solely contributed by QSOs irrespective of the \fesc\ value. Additionally at these energies, the intensity of the UVB depends on the average SED that is used to estimate the QSO emissivity. The QSO SEDs over large wavelength range can be approximated by a power law of the form $L_{\nu} \propto \nu^{\alpha}$ \citep[]{Francis91, VandenBerk01, Scott04, Shull12a}. Following \citet[]{Stevans14}, KS15 UVB uses $\alpha = -1.41$ at \lam<1000 \AA\ while HM12 uses $\alpha = -1.57$ \citep[]{Telfer02}. Note that, the power law has been observed only up to 2 Ryd \citep[i.e., for \lam >475 \AA,][]{Stevans14}, which is extrapolated to higher energies in UVB models. At  $\rm E > 4$~Ryd, the KS15 UVB has a factor of $\sim3$ higher intensity compared to HM12  due to the use of updated QSO emissivity and SED. The low ionized species (such as \OII, \OIII, \NII, \NIII, \CIII) have ionization energies $\rm E = 1-4$~Ryd.
At these energies, the UVB is contributed by radiation from QSOs as well as galaxies. The contribution of galaxies depends on the assumed value of \fesc. Since we are using the KS15 UVB  with \fesc=0\%, only QSOs contribute to UVB. Here, the KS15 UVB has a factor of 2 higher intensity due to the updated QSO emissivity. 

In Table~\ref{tab1}, we list the seven \NeVIII\ systems\footnote{We do not list the \NeVIII\ systems detected at \zabs = 0.927 towards PG~1206+549 by \citet[]{Tripp11}, as the authors do not provide the column density measurement of \OVI\ which is important in our analysis. Also, we do not list the \NeVIII\ absorption detected towards QSO J1154+4635 by \citet[]{Bordoloi16} as no information about the absorption redshift or of the low ionized species is provided which are also important in our analysis.} detected along the line-of-sight of five QSOs. The range of the absorption redshift covered by these systems is \zabs$ = 0.2-0.7$. In all these cases, \NeVIII\ absorption is co-aligned in velocity space with the \OVI\ absorption. Therefore, as is the case in most of the previous studies of \NeVIII\ absorbers, we will assume \OVI\ and \NeVIII\ to be co-spatial.
The column densities of \NeVIII\ and \OVI\ are well measured in these systems. Using previous UVBs, i.e., HM96, HM01 and HM05, photoionization models of most of the systems suggested a large line-of-sight thickness of >1~Mpc (as provided in Table~\ref{tab1}). At such large thickness, the width of the \NeVIII\ absorption due to the expansion of the universe will be much larger than the observed width, $b_{\neviii}$, which is in the range of 23 and 70 \kms, thereby ruling out a photoionized origin of \NeVIII. However there is only one system  where photoionization could explain the observed \NeVIII\ with the line-of-sight thickness measured being $\sim$186 kpc \citep[detected at \zabs=0.59961 towards PG~1407+265,][]{Hussain15}. 
Therefore, apart from this system, all the other \NeVIII\ systems are suggested to be originating from a collisionally ionized gas at temperatures $\log[T~(\rm K)] = 5.4-5.7$ with inferred line-of-sight thickness ($l$) $\lesssim 100$ kpc. At these high temperatures one expects the ionization fraction of \HI, \fhi $\sim 6\times10^{-7}$ in the collisional ionization models of \citet[]{Gnat07}. For $N_{H}$ in the range $10^{19}-10^{20}$ \cmsq, as suggested by the models, the expected $N(\HI)$ is $6\times(10^{12}-10^{13})$ \cmsq\ and the $b$-parameter of the absorption is in the range 66 and 93 \kms. This will produce a BLA associated with the \NeVIII\ absorption. Based on the presence of excess absorption in the wings of the \lya\ absorption, on top of the absorption profile predicted from the higher Lyman series lines, $N(\HI)$ associated with the \NeVIII\ is measured in the case of two systems: \zabs = 0.20701 towards HE0226--4110 \citep[]{Savage05a,Savage11} and \zabs = 0.32566 towards 3C 263 \citep[]{Narayanan09,Narayanan12}. For the remaining five absorbers, \lya\ absorption is not covered in the COS spectrum. For, \zabs=0.70152 and 0.72478 systems towards PG1148+549 \citep[]{Meiring13} and one of the \NeVIII\ components for the \zabs=0.59961 towards PG1407+265 \citep[]{Hussain15}, no \HI\ absorption is detected through \lyb\ or higher Lyman series lines. The lack of direct detection of BLAs in these cases suggests that if the gas is collisionally ionized than the metallicity has to be higher. 

Here, we re-analyse all these \NeVIII\ systems in the framework of photoionization by the KS15 UVB using \CLOUDY\ \citep[v13.03;][]{Ferland13}. We assume the absorbing gas to be plane-parallel slabs of constant density having relative abundances as observed in the solar photosphere \citep[]{Grevesse10}. To understand the ionization mechanisms of these absorbers, we first explore the total hydrogen density \nh\ and temperature $T$ range of the gas giving rise to the observed $N(\NeVIII)/N(\OVI)$. Constraints on \nh\ and metallicity $Z$\footnote{$Z$ is the linear metal abundance of the absorber with respect to solar photospheric abundances.} will give the line-of-sight thickness $l$, of the \NeVIII\ bearing gas cloud which is determined using the following equation: 
\begin{align}  
& l  =  \frac{N_{\neviii}}{n_{\rm \tiny H}f_{\neviii}} \frac{1}{Z} 
        \left(\frac{\rm Ne}{\rm H}\right)_{\odot}^{-1} & \label{eqn1}  \\ 
     & \simeq 200~{\rm kpc} \left(\frac{N_{\neviii}}{10^{14}}\right) 
     \left(\frac{n_{\rm \tiny H}}{10^{-5}}\right)^{-1}  
     \left(\frac{f_{\neviii}}{0.18}\right)^{-1} 
     \left(\frac{1}{Z}\right)\left(\frac{\rm Ne}{\rm H}\right)_{\odot}^{-1}.&  
\nonumber     
\end{align}    
Here, $f_{\neviii} \equiv n_{\neviii}/n_{\tiny \rm Ne}$ is the ionization fraction of \NeVIII\ (as computed by \CLOUDY) and $\left(\frac{\rm Ne}{\rm H}\right)_{\odot}$=$-4.07$ is the solar relative abundance of neon. Therefore, to determine $l$, we  need to obtain the metallicity of the gas associated with the \NeVIII\ phase. In the absence of $N(\HI)$ measurements from the \NeVIII\ phase, one assumes the gas phase metallicity of the \NeVIII\ phase to be similar to the low ionization phase. As we pointed out before, to estimate the metallicity of low ions, the assumed spectrum of UVB is crucial. 

 \begin{figure} 
  \centerline{
   \vbox{
   \centerline{\hbox{ 
   \includegraphics[totalheight=0.30\textheight, trim=0cm 0cm 0.75cm 0cm, clip=true]{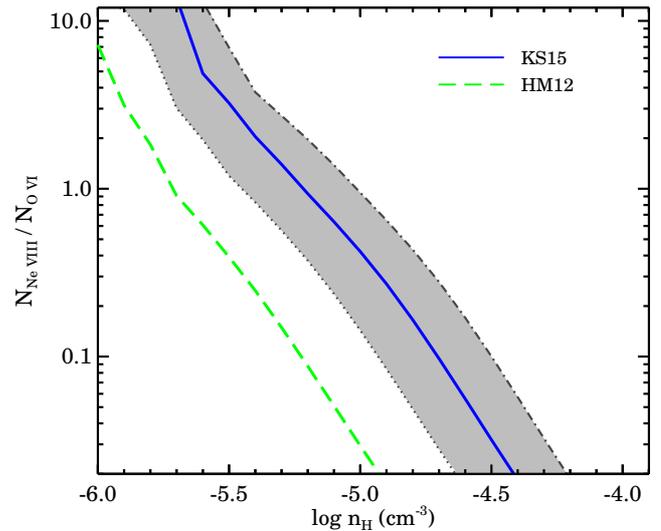}     
   }}
   }}  
 \caption{Results of photoionization model for \NeVIII\ and \OVI\ for an absorber at  $z$=0.2 using KS15 and HM12 UVB. Blue line represents the model predicted ratio for KS15 UVB while the dashed green line is for HM12 UVB. The gray shaded region shows the results of the photoionization model using KS15 UVBs generated with different QSO SED by changing $\alpha=-1.26$ (in dot-dashed curve) to $-1.56$ (in dotted curve) covering 1$\sigma$ range in $\alpha$ as measured by \citet[]{Stevans14}. For a given (\NeVIII/\OVI) ratio, \nh\ from KS15 model is a factor of $\sim$2-5 higher than that of HM12. This is due to the updated QSO emissivity used in KS15 UVB.}              
   \label{fig2}      
   \end{figure} 

\section{Ionization models for \NeVIII\ systems}\label{sec3}

As a demonstration of our approach, we run \CLOUDY\ models for $z$ = 0.2. Initially, to see the effect of UVB on the derived \nh, we compute the ratio of $N(\NeVIII)/N(\OVI)$ as a function of \nh\ under the photoionization equilibrium for KS15 and HM12 UVB. 
The results are shown in Fig.~\ref{fig2}. It is clear from this figure that for any observed value of $N(\NeVIII)/ N(\OVI)$, the \nh\ determined using KS15 UVB is a factor of $\sim3$ higher than what we find using HM12 UVB. Therefore, for a given metallicity and $N(\rm H)$, $l$ calculated for the KS15 UVB will be $3$ times smaller\footnote{Note that, the KS15 UVB uses the cloud distribution as provided in \citet[]{Haardt12}. We also model these \NeVIII\ systems using the UVB generated with \citet[]{Inoue14} cloud distribution and find the photoionization model results to be similar to KS15 UVB.} as compared to that derived using the HM12 UVB. There is a constant shift along the density axis between the two models as the spectral shape of the two UVBs are similar in the energy range of interest (see Fig.~\ref{fig2}).
As mentioned earlier, these high ions are affected by the UVB photons with $\rm E > 4$~Ryd where the contribution to UVB depends both on QSO emissivities as well as on the average SED of the QSOs. Therefore, we also run \CLOUDY\ models with KS15 UVB generated for $\alpha = -1.26$ to $-1.56$ consistent within 1$\sigma$ range in $\alpha$ ($= -1.41\pm0.15$) measured by \citet[]{Stevans14}. The resultant ratio of $N(\NeVIII)/N(\OVI)$ is shown in as gray-shaded region in Fig.~\ref{fig2}. Taking into account this uncertainty in SED, the measured \nh\ (and hence $l$) is a factor of $\sim 2-5$ higher (smaller) as compared to the one derived using the HM12 UVB. Note that, in all these models, the temperature ($T \sim 3-8\times10^{4}$ K) of the gas is self-consistently calculated by \CLOUDY\ which mainly arises due to photo-heating by the UVB. However, in the cosmological hydrodynamical simulations the gas temperature  at a given \nh\ can also be decided by other heating and cooling process. In order to account for this, we need to consider models over the temperature density plane from which \NeVIII\ systems can originate \citep[see for example,][]{Thor13}. 

\begin{figure*}
\centering
\includegraphics[totalheight=0.75\textheight, trim=7.5cm 0cm 0cm 0cm, clip=true, angle=90]{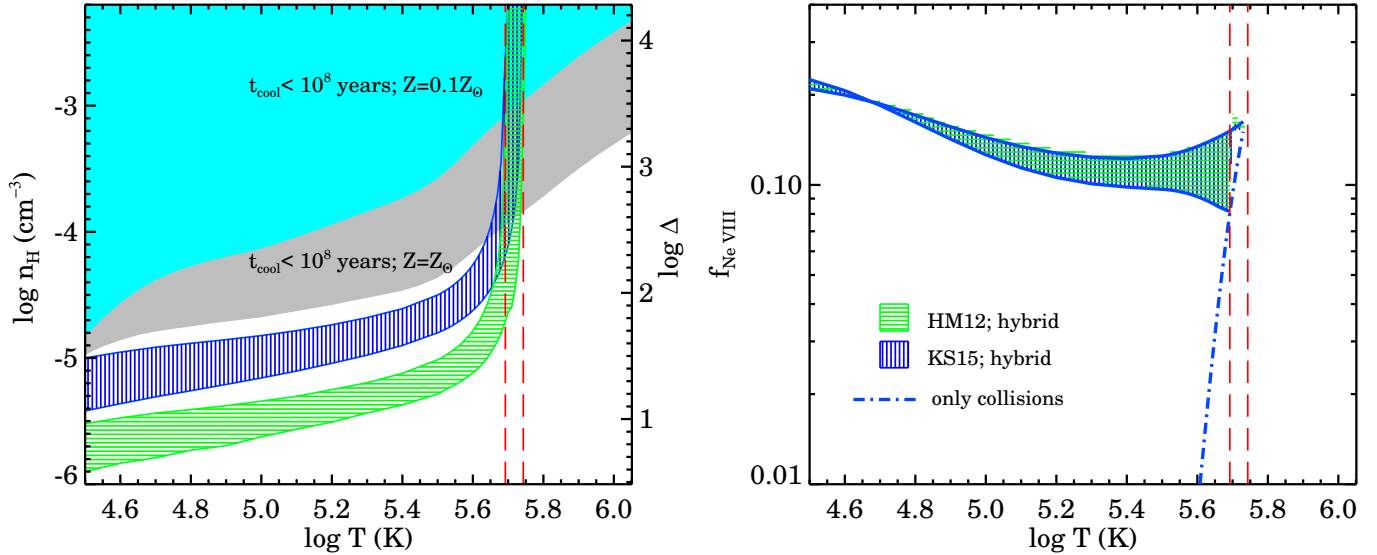}
\caption{\emph{Left panel:} The  \nh$-T$ plane showing the origin of observed \NeVIII\ systems at $z=0.2$ obtained to satisfy the observed ratio of \NeVIII\ to \OVI\ column densities. The hatched region with vertical blue lines and horizontal green lines show the \nh$-T$ plane from where \NeVIII\ systems can arise in the presence of KS15 and HM12 UVB, respectively. The gray and cyan shaded region shows \nh$-T$ plane where the cooling times scales to reach $T=10^4 K$ is less than 10$^8$ years for gas with $Z=Z_{\odot}$ and $Z=0.1Z_{\odot}$, respectively. Red dash line shows the region dominated by collisions. We also show the overdensities ($\Delta$) measured corresponding to each \nh. \emph{Right panel:} The ionization fraction of \NeVIII, $f_{\neviii}$, with temperature from hybrid models of left panel for KS15 and HM12 UVB at $z=0.2$. The \nh\ at each $T$ is taken from the left panel for the respective UVB model. The dot-dash line shows $f_{\neviii}$ only from collisions. Red dash lines are same as in the left panel.}
\label{fig3}
\end{figure*}

\begin{figure} 
\centerline{
\vbox{
\centerline{\hbox{   
\includegraphics[width=0.59\textwidth, clip=true, angle=00]{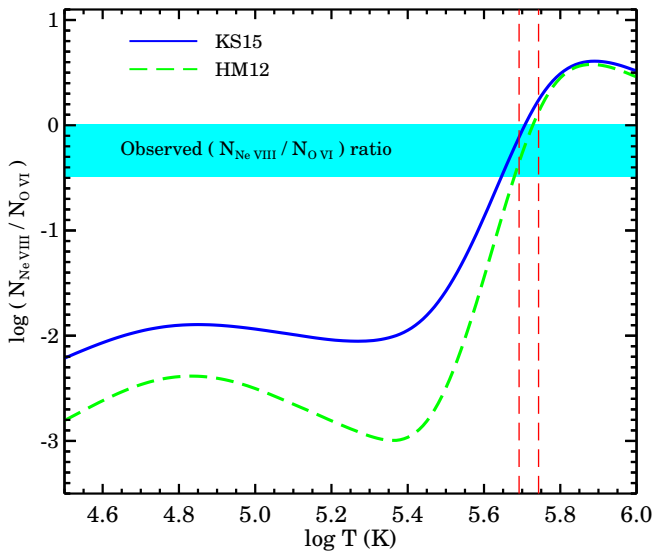}
}}
}}  
\caption{A ``hybrid'' photoionization model computed for $z$=0.2 assuming the gas to be isochoric i.e., having a constant density. We choose a constant density \nh=$10^{-4}$~\cmcb\ as it is the critical density above which the (\NeVIII/\OVI) ratio can be produced with collisional ionization. The green and blue curves represent the model solutions for the HM12 and KS15 UVB respectively. The shaded region represent the observed column density ratio of \NeVIII/\OVI\ while the red dotted lines show the region dominated by collisions. At low temperatures the model predicted $N(\NeVIII)/N(\OVI)$ is at least an order of magnitude lower than the observed ratio.}
\label{fig4}              
\end{figure} 

We ran several \CLOUDY\ models in the presence of different UVB at $z=0.2$ on a grid of constant temperatures and \nh\ to reproduce the observed range in the column density ratios of \NeVIII\ to \OVI\ (i.e. 0.33 to 1.02, see Table~\ref{tab1}) and the results are shown in Fig.~\ref{fig3}. We refer to these models as ``hybrid models''. 
In the left-hand panel of Fig.~\ref{fig3}, the hatched region with vertical blue lines and horizontal green lines show the allowed range in the \nh$-T$ plane from where $N(\NeVIII)/N(\OVI)$ constraints can be obtained in the case of KS15 and HM12 UVB, respectively. Irrespective of the UVB model, the required \nh\ increases slowly with increasing gas temperature up to a certain temperature and then it rapidly diverges after a critical density (i.e., \nh$\sim2\times10^{-5}$~\cmcb\ for HM12 and $\sim5\times10^{-5}$~\cmcb\ for KS15). The region of \nh$-T$ plane where \nh\ increases slowly is dominated by photoionization. This slow increase is mainly due to the temperature dependence of the  recombination coefficients of several ions. The small region in the  \nh$-T$ plane where \nh\ increases rapidly is dominated by collisions. To show the relative contribution of collisions to the \NeVIII\ production, in the right-hand panel of Fig.~\ref{fig3}, we show the ionization fraction of \NeVIII, $f_{\neviii}$, as a function of temperature. At each temperature, the \nh\ is taken from  \nh$-T$ plane for the respective UVB model as shown in the left panel of Fig.~\ref{fig3}. The $f_{\neviii}$ from collision peaks sharply at a very small temperature range. It is comparable to the  $f_{\neviii}$ obtained in the hybrid model with $5.68<\log T(\rm K)<5.74$ as shown in the left-hand panel of Fig.~\ref{fig3}. 

Given the high densities inferred for the gas to be in collisional ionization equilibrium in the presence of UVB (i.e., the baryonic overdensity $\Delta>100$, see the left hand panel in Fig.~\ref{fig3}), the gas at the above mentioned range in temperature can cool very efficiently. 
We estimate the cooling time, $t_{\rm cool}$, the time required by the gas in \nh$-T$ plane to reach $T=10^4$K, using the cooling curves of \citet[]{Schure09}. The gray and cyan shaded region in the left-hand panel of Fig.~\ref{fig3} show the region of \nh$-T$ plane that can cool in $t_{\rm cool} < 10^8$ years, for a gas with metallicity $Z_{\odot}$ and $0.1Z_{\odot}$, respectively. It suggests that most of the collisionally ionized \NeVIII\ systems can cool very fast to the temperatures where \NeVIII\ can originate from photoionization. Now the cooling can be isochoric (constant density) or isobaric (constant pressure): i.e., increasing density with decreasing temperature. We consider an isochoric cooling gas (i.e., a gas with a constant \nh$=10^{-4}$~\cmcb, which is the critical density above which the $N(\NeVIII)/N(\OVI)$ ratio can be produced by collisional ionization) in the presence of UVB in Fig.~\ref{fig4}. The model predicted $N(\NeVIII)/N(\OVI)$ ratio at low temperatures is at least an order of magnitude lower than the observed ratio. The situation will be worse in the case of isobaric where one expects the density to be higher at low $T$. Therefore, in order for the collisionally ionized gas to reproduce the observed $N(\NeVIII)$ and $N(\NeVIII)/N(\OVI)$ ratio, one needs to avoid the rapid cooling. This rapid cooling issue can be avoided if we have (i) the gas metallicity to be low (i.e., $Z<0.1Z_{\odot}$) or (ii) the cloudlets containing this gas are being continuously pumped by mechanical energy from nearby sources at a rate that can compensate the inferred cooling rate.  

In the left-hand panel of Fig.~\ref{fig3}, we also show the overdensities ($\Delta$) corresponding to each \nh\ (from \nh$=1.9\times10^{-7}$(\cmcb)$(1+z)^{3}\Delta$). Large overdensities ($\Delta > 300$ for KS15 UVB; $\Delta > 80$ for HM12 UVB) are required to maintain the collisional ionization equilibrium within $5.68<\log T\rm(K)<5.74$ (where collisions dominate). Such overdensities reveal that the absorbers should originate in a dense galactic environment or from hot halos. This is consistent with the detection of a disk galaxy in the vicinity of the \NeVIII\ absorber towards HE~0226--4110 \citep[]{Mulchaey09} with luminosity $L$=$0.25L_{\star}$, 
at an impact parameter $\rho$=109 kpc. Here, $L_{\star}$ is the characteristic galaxy luminosity. Other than the above absorber, three absorbers at \zabs = 0.49509 towards PKS~0405--123, 0.72478 towards PG~1148+549 and 0.59961 towards PG~1407+265 are also found to be associated with L=0.08$L_{\star}$, L=$L_{\star}$ and a sub-$L_{\star}$ galaxy at an impact parameter of 110 kpc \citep[]{Chen09}, 217 kpc \citep[]{Meiring13} and 89 kpc (J. Werk, \emph{private communication}) respectively, in their vicinities. But there is no direct detection of galaxies, although the measured metallicities are solar, for the absorber at \zabs = 0.32566 towards 3C~263 \citep[]{Narayanan09, Narayanan12}.

\begin{figure*}
\centering
\includegraphics[totalheight=0.75\textheight, trim=7.5cm 0cm 0cm 0cm, clip=true, angle=90]{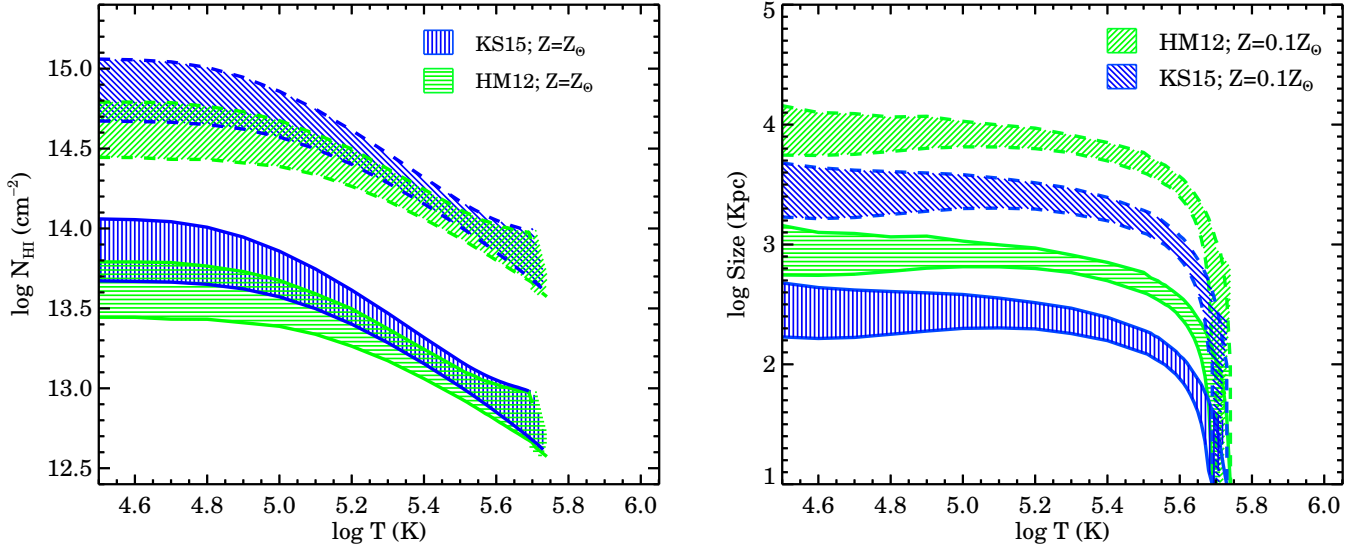}
\caption{\emph{Left panel:} The $N(\HI)$ associated with \NeVIII\ system at $z=0.2$ at different constant temperatures. The blue and green hatched regions are the $N(\HI)$ in the presence of KS15 and HM12 UVB, respectively. Here, we assumed $N(\NeVIII)=10^{14}$ \cmsq\ consistent with the observations. Results are shown for the gas with $Z=Z_{\odot}$ and $Z=0.1Z_{\odot}$ respectively. \emph{Right panel:} The line-of-sight thickness of \NeVIII\ systems with temperature at $z=0.2$ in the presence of KS15 and HM12 UVB. For each model, the $n_{\rm H}$ is taken from the \nh$-T$ plane  of Fig.~\ref{fig3} (left-hand panel).}
\label{fig5}
\end{figure*}

We also calculate the $N(\HI)$ associated to the \NeVIII\ phase in these models. For the illustrative purpose, we take $N(\NeVIII)=10^{14}$ \cmsq, consistent with the observations (see Table~\ref{tab1}). The results are shown in the left-hand panel of Fig.~\ref{fig5} for a gas with the solar metallicity ($Z_{\odot}$) and $0.1Z_{\odot}$. To get the observed $N(\HI)$ (i.e., $\log N(\HI)\sim14$) from the photoionized gas, the metallicity should be of the order of $Z_{\odot}$ and more. However, the collisionally ionized gas ($T > 10^{5.5}$ K) can have low metallicities ($0.1Z_{\odot} < Z < Z_{\odot}$). We have also calculated the line-of-sight thickness for these systems, as shown in the right-hand panel of Fig.~\ref{fig5}. For the gas in photoionization equilibrium with the KS15 UVB model, the gas with metallicity $Z \geqslant Z_{\odot}$ can come from clouds with thickness $l\leqslant 400$ kpc. The thickness of the low metallicity clouds originating due to photoionization is too large to be physical. These systems are better explained with collisional ionizations since the gas with temperatures where collisions dominate, can come from very small clouds (as can be seen in the right-hand panel of Fig.~\ref{fig5}).  However it is interesting to investigate the environment of such systems to study the mechanism keeping them at these temperatures. Moreover to understand the ionization scenarios of \NeVIII\ absorbers it is important to measure the metallicity and $N(\HI)$ of the gas.

\section{Re-analysis of the \NeVIII\ absorbers}\label{sec4}

Here we re-analyse the \NeVIII\ absorbers identified in Table~\ref{tab1} in the framework of photoionization using the KS15 UVB. For a few systems, \NeVIII\ absorption is seen in multiple components. For the sake of simplicity, in this work, we model the \NeVIII\ component at \vrel=0~\kms, unless otherwise specified. Owing to the non-detection of $N(\HI)$ associated with the \NeVIII\ phase, we assume the gas phase metallicity of this \NeVIII\ phase to be similar to what we derive from the low ionization phase. We use the column density ratios of \OIII/\OIV, \OIV/\OIII\ and/or \NII/\NIII\ to constrain the low ionization gas phase density and hence measure the metallicity of this phase. We use this derived metallicity, assuming it to be similar to the metallicity of the high ionization phase, as an input in our photoionization model for \NeVIII\ phase. The density of the \NeVIII\ phase is constrained using the column density ratio of \NeVIII/\OVI. Below we briefly discuss photoionization model of each of the \NeVIII\ systems.   
\begin{table*}
\caption{Re-analysis of \NeVIII\ absorbers under photoionization equilibrium using the KS15 UVB}
\label{tab2}
\begin{center}
\begin{tabular}{lcccccccc} \hline \hline 
  &   & Ion ratios used &   &   &   &   &  & \\
QSO & \zabs & to constrain low ion & $[$X/H$]^{a}$ & $\log$~[\nh]$^{b}$ & $\log [N(\HI)]^{c}$ & $\log [N_{\rm H}$] & $l^{d}$ \\
  &   & gas density \nh &   & (\cmcb) & (\cmsq) & (\cmsq) & (kpc) \\ \cmidrule(){1-1} \cmidrule(lr){2-2} \cmidrule(lr){3-3} \cmidrule(lr){4-4}  \cmidrule(lr){5-5} \cmidrule(lr){6-6} \cmidrule(lr){7-7} \cmidrule(lr){8-8} 
HE~0226--4110 & 0.20701 & \OIII/\OIV & 0.12 & $-4.90$ & 13.90 & 18.56 & 94  \\
3C~263 & 0.32566 & \OIII/\OIV       & 0.00  & $-5.10$ & 13.59 & 18.86 & 299  \\
PKS~0405--123 & 0.49509 & \OIII/\OIV & 0.46 & $-4.50$ & 13.69 & 18.14 & 22  \\
PG~1148+549 & 0.68381 & \OIV/\OIII  & 0.32  & $-4.40$ & 13.58  & 18.40  & 20    \\
PG~1148+549 & 0.70152 & \OIV/\OIII  & >0.95 & $-4.40$ & <12.86  & <17.64  & 4   \\
PG~1148+549 & 0.72478 & \OIV/\OIII  & >0.79 & $-4.70$ & <12.54  & <17.88  & 12  \\
PG~1407+265 & 0.59961 &  \NII/\NIII & >0.27$^{e}$ & $-4.80$ & <12.90  & <18.27  & 38    \\ \hline 
\end{tabular}
\begin{flushleft}
{\bf Notes:} \\
$^{a}$~Logarithmic metallicity with respect to solar photospheric abundances. The metallicities listed in the table for the high ions are measured from the low ion photoionization model, assuming them to be similar to the high ionization phase. The lower limits in metallicities are because of the non-detection of higher \HI\ Lyman series absorption lines associated with the \NeVIII\ absorber. \\
$^{b}$~Total hydrogen density, \nh, associated with the photoionized \NeVIII\ gas phase as measured from the photoionization model. \\  
$^{c}$ The neutral hydrogen column density associated with the photoionized \NeVIII\ gas phase.\\
$^{d}$~The line-of-sight thickness for each of the photoionized \NeVIII\ absorber is measured using Eq.~\ref{eqn1}.\\
$^{e}$ The metallicity measured for this absorber is based on oxygen (\OII).

\end{flushleft}
\end{center}
\end{table*}


\subsection{HE 0226--4110 \citep[]{Savage05a,Savage11}}\label{z02}
A \NeVIII\ absorber is detected along the line-of-sight of QSO HE 0226--4110 at \zabs=0.20701. Along with \NeVIII, a BLA is also detected having $\log [N(\HI)$(\cmsq)$]$=$13.87\pm0.08$ and $b$=$72^{+13}_{-6}$~\kms. For this absorber, using HM96 UVB, the authors ruled out a possible origin of \NeVIII\ under photoionization equilibrium. Assuming the low ion gas phase metallicity (i.e., $[$X/H$]$=$-0.5$) to be similar to the \NeVIII\ phase, their photoionization model solutions for \NeVIII\ are: \nh$=4.5\times10^{-7}$ \cmcb\ and line-of-sight thickness $l \sim 11$~Mpc.   

Here, using the column density ratio $N(\OIII)/N(\OIV)$, we constrain the low ionization gas phase density and hence determine the metallicity of this phase to be $[$X/H$]=0.12$. We use this measured metallicity in photoionization models for \NeVIII\ phase and the model solutions  are listed in Table~\ref{tab2}. With the updated UVB, we find the \nh\ to be $\sim$2 orders of magnitude more than what was derived earlier (using HM96 UVB). As a result, we obtain $\sim$2 orders of magnitude less line-of-sight thickness ($l$=94~kpc) using KS15 UVB. Thus, owing to the small line-of-sight thickness measured, this \NeVIII\ absorber can have a photoionized origin. 

However, if we assume, that \HI\ column density associated with the \NeVIII\ to be a BLA as suggested by \citet[]{Savage11}, then from collisional ionization equilibrium models of \citet[]{Gnat07}, the metallicity of the \NeVIII\ phase turns out to be $[$X/H$]$=$-0.94$ \citep[is within errors of what have been reported by][]{Savage11}. This metallicity is a factor of $\sim$12 less compared to that derived from photoionization model of low ions. From the discussions presented in Section~\ref{sec3}, it is clear that the low metallicity gas has larger cooling timescales and sustain the \NeVIII\ phase at a detectable level for a longer period of time. In summary, photoionization is a viable option for this system if the \NeVIII\ phase has the same metallicity as the low ion phase. However, if the BLA identified by \citet[]{Savage11} is correct, then \NeVIII\ can be consistently produced by collisions with the absorbing gas having lower metallicity than that of the gas producing low ion absorption. 

\subsection{3C 263 \citep[]{Narayanan09,Narayanan12}}
Along with \NeVIII\ absorption at \zabs=0.32566 towards the line-of-sight of QSO 3C 263, a Broad \lya\ is also marginally detected with $\log [N(\HI)$(\cmsq)$] \sim 13.20$ and $b=93$~\kms. For this absorber, from low ion photoionization model using HM96 UVB, the metallicity reported  by \citet[]{Narayanan09} is $[$X/H$]$=$-0.38$. Assuming this metallicity to be similar to the \NeVIII\ phase, the photoionization model solutions for the \NeVIII\ phase are: \nh$\sim5.0\times10^{-6}$~\cmcb, $N_{\rm H}\sim10^{20}$~\cmsq\ and line-of-sight thickness $\sim5.6$~Mpc. Owing to the large thickness measured, the authors ruled out a photoionized origin for \NeVIII. 

Here, like above, by using the column density ratio of $N(\OIII)/N(\OIV)$ as a constraint on low ionization gas density, we obtain the metallicity of the low ionization phase to be solar (i.e., $[$X/H$]=0.00$). We use this measured metallicity in photoionization models for \NeVIII\ phase and the results are shown in Table~\ref{tab2}. We find the \nh\ measured is a factor $\sim$2 more than what was derived using HM96 UVB. As compared to HM96, we obtain much smaller line-of-sight thickness ($l=299$~kpc) ensuring the \NeVIII\ absorber can have a photoionized origin.

If the BLA is associated with the warm \NeVIII\ phase, then under the collisional ionization models of \citet[]{Gnat07}, we measure the metallicity to be $[$X/H$]$=$-0.53$. Like the above system, this absorber too can be produced via collisions provided we have gas metallicity of the high ion phase to be less than that of the low ion phase by a factor $\sim$3.

\subsection{PKS 0405--123 \citep[]{Narayanan11}}

A \NeVIII\ absorber is detected at \zabs=0.495096 towards QSO PKS 0405--123. Using the HM01 UVB and solar metallicity in their photoionization models\footnote{The photoionization model for the low ions detected in this absorber are discussed in detail by \citet[]{Howk09} where the observations were carried out by the {\it FUSE}. Using HM01 UVB and constraining the ionization parameter using the column density ratios of $N(\OIV)/(\CIII)$, $N(\OIII)/(\OIV)$ and $N(\OIV)/(\OVI)$, the metallicity reported for the low ionized phase is in the range $-0.62\leqslant [$X/H$] \leqslant -0.15$.}, the authors reported that photoionization is not responsible for the production  of \NeVIII\ in this absorber. Their photoionization model solutions for \NeVIII\ are: \nh\ $\sim8\times10^{-6}$~\cmcb, $N_{H}\sim6\times10^{19}$~\cmsq\ and line-of-sight thickness is $2.5$~Mpc. 

Using the KS15 UVB, the results of the photoionization model for the \NeVIII\ phase of this absorber are shown in Table~\ref{tab2}. We obtain the metallicity of the low ionization (and hence high ionization) gas phase to be $[$X/H$]= 0.46$. In addition, \nh\ that is required to reproduce the observed $N(\NeVIII)/N(\OVI)$ is a  factor of $\sim$4 higher as compared to what was derived using the HM96 UVB. Also, with the small measured line-of-sight thickness ($l=22$~kpc), this absorber too can have a photoionized origin.

\subsection{PG 1148+549 \citep[]{Meiring13}}
Three \NeVIII\ absorbers at \zabs=0.68381, 0.70152 and 0.72478 respectively were observed towards the line-of-sight of the QSO PG1148+549. Here the authors using the HM96 UVB, concluded that photoionization models cannot explain the \NeVIII\ absorption due to their measured unrealistically large cloud sizes (>1~Mpc). For \zabs=0.68381, the metallicity reported by \citet[]{Meiring13} is $[$X/H$]$=$-0.5$, while for \zabs=0.70152 and 0.72478, $[$X/H$]$>~$0.2$ and $[$X/H$]$>~$0.0$ respectively, based on photoionization model solutions for the low ions (\CIII, \OIII, \OIV, \NIV) using HM96 UVB. The lower limits in metallicities for the two systems (\zabs=0.70152, 0.72478) are due to the non-detection of higher \HI\ Lyman series absorption lines. 

Using the KS15 UVB and constraining the low ionization gas phase density with the column density ratio of \OIV/\OIII, our photoionization model measured the low ionization (and hence the high ionization) gas phase metallicities to be solar to supersolar. In Table~\ref{tab2}, we summarize the photoionization model results of the \NeVIII\ phase associated with the three absorbers. With the high metallicity measured, as discussed in the previous section, we expect the cooling timescales to be much shorter in this case if the gas is in collisional equilibrium. Thus it is natural to conclude that \NeVIII\ is originating from a photoionized gas. The derived line-of-sight thickness ($\leqslant20$~kpc) also are not unrealistic to rule out the photoionization models.

\subsection{PG 1407+265 \citep[]{Hussain15}}
\NeVIII\ absorption is detected in four components (\vrel=$-150,75,0,60$~\kms) towards the line-of-sight of QSO PG 1407+265 at \zabs=0.59961. Using the HM05 UVB and solar metallicity, the authors could explain the \NeVIII\ absorption component at \vrel=$-$150~\kms. under photoionization equilibrium models. Their photoionization model solutions for this \NeVIII\ absorber are \nh$=5.0\times10^{-6}$~\cmcb, $\log N_{\rm H}=19.4$~\cmsq\ and line-of-sight thickness is $l$=186~kpc. 

We too, in this work, using the KS15 UVB, also could explain the \NeVIII\ absorber at \vrel=$-150$~\kms\ under photoionization. The  photoionization model results for the \NeVIII\ phase are listed in Table~\ref{tab2}. Like the systems discussed above, the metallicity of the low ions (and hence of the high ions) for this absorber are measured to be supersolar. Here the \nh\ is a factor $\sim3$ higher as compared to what have been reported by \citet[][]{Hussain15} from their photoionization model. With even small line-of-sight thickness ($l=38$~kpc) measured with the updated UVB, photoionization too proves to be the natural explanation for this absorber.  
\subsection{Discussions}

The line-of-sight thickness $l\geqslant100$~kpc measured for the two photoionized absorbers: \zabs=0.20701 towards HE~0226--4110 and \zabs=0.32566 towards 3C~263, as discussed above, are still large. However, at low-$z$, such large cloud sizes are comparable to the sizes of oxygen-rich halos of isolated star-forming galaxies \citep[see][]{Tumlinson11, Werk13, Werk14} and of the CGM traced by \OVI\ absorbers \citep[see][]{Muzahid14}.

Given the supersolar metallicities measured in most of the photoionized \NeVIII\ absorbers, it is interesting to see whether normal galaxies could enrich such \NeVIII\ bearing gas clouds. \citet[]{Peeples14}, accounting for the total mass of metals produced by $z\sim0$ galaxies, estimated that 20--25\% of total metal mass produced remains in galaxies i.e., in stars, interstellar gas and interstellar dust. The remaining 75--80\% can be distributed in the outskirts of galaxies. Using this fact, we can estimate the stellar mass of galaxies required to enrich \NeVIII\ bearing clouds. For that, we use the relation between the total mass of metal oxygen produced by galaxies (via a Type II supernovae) and stellar mass of galaxies \citep[see Equation~2 of][]{Peeples14}. Under the approximation that these absorbers have a spherical geometry, the total mass of metal oxygen, $M_{\rm oxy}$, in the absorbers is 
$M_{\rm oxy} \simeq 10^{5}~M_{\odot} \left (\frac{n_{\rm \tiny H}}{\rm cm^{-3}}\right) \left(\frac{l}{\rm kpc}\right)^{3}Z$. Here $l$, \nh, and $Z$ are respectively, the line-of-sight thickness, total hydrogen density, and metallicity of the absorber. 

Assuming that these \NeVIII\ absorbers are probing less than 5\% (out of 75\%) of metals seen in the outskirts of galaxies, the stellar mass (in terms of oxygen mass) in these galaxies can be written as
\begin{equation*}%
   \log(M_{\star}/M_{\odot})  < [ \log(M_{\rm oxy}/(0.05\times0.75)) + 1.71481 ] / 1.0146 
\end{equation*}
We find the stellar masses to be $ < 2\times10^{10} M_{\odot}$, even for the largest cloud with thickness of 299 kpc. We convert the stellar masses required for each absorber into luminosities using equation~7 of \citet[]{Paranjape15}, where we use the typical mass to light ratio of 0.5384 in SDSS $r-$band (with ($g-r)=0.4$) obtained for the blue star-forming galaxies with stellar masses $<10^{10}~M_{\odot}$ \citep[see Figure~4,][]{Paranjape15}.
By comparing these luminosities with the observed luminosity function of galaxies as described in \citet[]{Ilbert05}, we find that all these associated galaxies can be sub-$L_{\star}$ galaxies. This is consistent with the detection of sub-$L_{\star}$ to $L_{\star}$ galaxies associated with some of the absorbers as discussed in Section~\ref{sec3}. Thus, the inferred high metallicities and large cloud thicknesses can be accounted for with only $\sim 5\%$ or less mass fraction of total metals produced by sub-$L_{\star}$ galaxies. While this back of the envelope calculation provides a sanity check, detailed understanding of the enrichment and spatial distribution of neon in galaxy halos requires more detections of \NeVIII\ absorbers and associated galaxy identifications. 
%


\section{Summary}\label{sec5}

Using the updated ultraviolet background KS15 at low-$z$, we re-analyse the intervening \NeVIII\ absorbers to understand the ionization mechanisms of the \NeVIII\ systems. The important conclusions are summarized as follows:

1. Using the updated KS15 UVB, \NeVIII\ and \OVI\ are simultaneously reproduced in a photoionized gas having total hydrogen densities of $n_{\rm \tiny H} = 4\times10^{-5} - 8\times10^{-6}$~\cmcb. This, in comparison to HM12, is a factor $\sim3$ higher. Also, except for a few, all these photoionized absorbers require near solar to supersolar ($Z = 1-3~Z_{\odot}$) metallicities. The line-of-sight thickness measured for these \NeVIII\ systems with KS15 UVB is $l<100$~kpc (except for one absorber with $l=299$~kpc) favoring a photoionized origin for all these absorbers. This is unlike, what have been found from previous studies of the \NeVIII\ absorbers \citep[except][]{Hussain15} with HM96, HM01 and HM05 UVB, where photoionization was ruled out purely based on large line-of-sight thickness ($> 1$ Mpc) of the cloud.

2. We show that the collisionally ionized gas with solar metallicity and having density above critical density can cool within $10^{8}$ years. This will make the detectability of the collisionally ionized \NeVIII\ phase difficult, unless there is a constant heating source present in the system. The cooling issue can be sorted out if the gas phase metallicity is lower. In a couple of cases, BLAs associated with the \NeVIII\ absorbers have been reported. The inferred $N(\HI)$ is consistent with collisional ionization of \NeVIII\ gas when the gas phase metallicity is less than that of the low ion phase. In such cases collisions are possible. Therefore, accurate determination or constraints of Broad \lya\ absorption associated with the \NeVIII\ phase is important to distinguish between photoionization and collisional ionization equilibrium models. Mere line-of-sight thickness arguments cannot rule out the photoionized origin of the \NeVIII\ phase. 

3. Photoionized with the KS15 UVB, \NeVIII\ absorbers probe $\lesssim 0.5\%$ of the cosmic baryon density as compared to what have been reported for collisionally ionized \NeVIII\ gas ($\approx 4\%$). This is because these photoionized absorbers have 10 times less total hydrogen column densities $\log N_{H}= 17.5-19.0$~\cmsq~ as compared to what one obtain from collisional ionization models.

\section{Acknowledgment}
We thank the anonymous referee for useful comments. TH acknowledges IUCAA for hospitality and support from Startup Research Grant YSS/2014/000338 (PI: N. Gupta) during the period of this work. 
\bibliographystyle{mnras}

\bibliography{thbib}

\def\aj{AJ}%
\def\actaa{Acta Astron.}%
\def\araa{ARA\&A}%
\def\apj{ApJ}%
\def\apjl{ApJ}%
\def\apjs{ApJS}%
\def\ao{Appl.~Opt.}%
\def\apss{Ap\&SS}%
\def\aap{A\&A}%
\def\aapr{A\&A~Rev.}%
\def\aaps{A\&AS}%
\def\azh{AZh}%
\def\baas{BAAS}%
\def\bac{Bull. astr. Inst. Czechosl.}%
\def\caa{Chinese Astron. Astrophys.}%
\def\cjaa{Chinese J. Astron. Astrophys.}%
\def\icarus{Icarus}%
\def\jcap{J. Cosmology Astropart. Phys.}%
\def\jrasc{JRASC}%
\def\mnras{MNRAS}%
\def\memras{MmRAS}%
\def\na{New A}%
\def\nar{New A Rev.}%
\def\pasa{PASA}%
\def\pra{Phys.~Rev.~A}%
\def\prb{Phys.~Rev.~B}%
\def\prc{Phys.~Rev.~C}%
\def\prd{Phys.~Rev.~D}%
\def\pre{Phys.~Rev.~E}%
\def\prl{Phys.~Rev.~Lett.}%
\def\pasp{PASP}%
\def\pasj{PASJ}%
\def\qjras{QJRAS}%
\def\rmxaa{Rev. Mexicana Astron. Astrofis.}%
\def\skytel{S\&T}%
\def\solphys{Sol.~Phys.}%
\def\sovast{Soviet~Ast.}%
\def\ssr{Space~Sci.~Rev.}%
\def\zap{ZAp}%
\def\nat{Nature}%
\def\iaucirc{IAU~Circ.}%
\def\aplett{Astrophys.~Lett.}%
\def\apspr{Astrophys.~Space~Phys.~Res.}%
\def\bain{Bull.~Astron.~Inst.~Netherlands}%
\def\fcp{Fund.~Cosmic~Phys.}%
\def\gca{Geochim.~Cosmochim.~Acta}%
\def\grl{Geophys.~Res.~Lett.}%
\def\jcp{J.~Chem.~Phys.}%
\def\jgr{J.~Geophys.~Res.}%
\def\jqsrt{J.~Quant.~Spec.~Radiat.~Transf.}%
\def\memsai{Mem.~Soc.~Astron.~Italiana}%
\def\nphysa{Nucl.~Phys.~A}%
\def\physrep{Phys.~Rep.}%
\def\physscr{Phys.~Scr}%
\def\planss{Planet.~Space~Sci.}%
\def\procspie{Proc.~SPIE}%
\let\astap=\aap
\let\apjlett=\apjl
\let\apjsupp=\apjs
\let\applopt=\ao


\label{lastpage}

\end{document}